\documentclass[]{article}
\newfont{\bb}{msbm10}
\def\Bbb#1{\hbox{\bb#1}}

\begin{document}

\author{B.G. Konopelchenko and G. Landolfi \and \textit{Dipartimento di Fisica,
Universit\`{a} di Lecce, 73100 Lecce, Italy.\thanks{%
Supported in part by PRIN 97 ``Sintesi''}}}
\title{On rigid string instantons in four dimensions }
\date{}
\maketitle

\begin{abstract}
Generalized Weierstrass formulae for surfaces in four-dimensional space $%
\Bbb{R}^{4}$ are used to study (anti)self-dual rigid string configurations.
It is shown that such configurations are given by superminimal immersions
into $\Bbb{R}^{4}$. Explicit formulae for generic (anti)instantons are
presented. Particular classes of surfaces are also analyzed.
\end{abstract}

String theory based on the combined Nambu-Goto-Polyakov action has been
intensively studied during the last decade (see \emph{e.g.} [\ref{re1},\ref
{re2}]). While formulated for the target space of arbitrary dimensions, it
exhibits special interesting properties in four dimensions. In this case, in
particular, one can introduce an additional topological ``$\theta $-term''
which modifies essentially the properties of the strings [\ref{re3},\ref{re4}%
]. Such a total action is of the form 
\begin{equation}
S=S_{N-G}+S_{P}+i\,\theta \,I  \label{e1}
\end{equation}
where 
\begin{eqnarray}
S_{N-G} &=&\mu _{0}\int [d\Sigma ]\quad \quad ,  \label{e2} \\
S_{P} &=&\frac{1}{\alpha _{0}}\int \overrightarrow{H}^{2}[d\Sigma ]
\label{e3}
\end{eqnarray}
and $I$ is the self-intersection number given by 
\begin{equation}
I=-\frac{1}{16\pi }\int [d\Sigma ]\,g^{ab}\partial _{a}t^{\mu \nu }\partial
_{b}t^{*\mu \nu }\quad .  \label{4}
\end{equation}
Here $X^{\mu }$ ($\mu =1,2,3,4$) denote coordinates of a surface in the
four-dimensional Euclidean space $\Bbb{R}^{4}$, $\xi _{a}\,$($a=1,2$) are
local coordinates on a surface, $g_{ab}=\frac{\partial X^{\mu }}{\partial
\xi ^{a}}\frac{\partial X^{\mu }}{\partial \xi ^{b}}$ ($a=1,2$) is the
induced metric, $[d\Sigma ]=\sqrt{\det g}\,d\xi _{1}d\xi _{2}$, $%
\overrightarrow{H}$ denotes a mean curvature vector, $t^{\mu \nu }=\frac{%
\varepsilon ^{ab}}{\sqrt{\det g}}\frac{\partial X^{\mu }}{\partial \xi ^{a}}%
\frac{\partial X^{\nu }}{\partial \xi ^{b}}$ and $t^{*\mu \nu }=\frac{1}{2}%
\varepsilon ^{\mu \nu \rho \sigma }t^{\rho \sigma }$.

Special classical configurations, called instantons, play an important role
in the string theory governed by the action (\ref{e1}). They are the
solutions of the system [\ref{re3}] 
\begin{equation}
\partial _{a}t^{\mu \nu }=\pm \partial _{a}t^{*\mu \nu }\,\quad .  \label{e5}
\end{equation}
Solutions of the system (\ref{e5}) and their properties have been studied by
different methods in a number of papers [\ref{re1}-\ref{re10}]. But the
complete description of instantons is still missing.

In this letter we use the generalized Weierstrass representation (GWR) for
surfaces in $\Bbb{R}^{4}$ to analyze solutions of the system (\ref{e5}). We
show that general instantons of the theory (\ref{e1}) are given by
superminimal immersions into $\Bbb{R}^{4}$. Explicit form of generic
instantons is presented. We consider also some special classes of surfaces
(developable surfaces, surfaces with flat normal bundle)\ which are of
non-instanton type but are of importance for the string theory too.

The generalized Weierstrass representation (GWR) for surfaces in $\Bbb{R}%
^{4} $ has been proposed recently in [\ref{re11},\ref{re12}]. It starts with
the two two-dimensional Dirac equations 
\begin{equation}
\begin{array}{l}
\psi _{1z}=p\varphi _{1}\quad \quad , \\ 
\varphi _{1\overline{z}}=-\overline{p}\psi _{1}\quad \,\,,
\end{array}
\qquad 
\begin{array}{l}
\psi _{2z}=\overline{p}\varphi _{2}\quad \quad , \\ 
\varphi _{2\overline{z}}=-p\psi _{2}\qquad .
\end{array}
\label{e6}
\end{equation}
where $z$ is a complex variable, bar denotes the complex conjugation, $\psi
_{\alpha },\varphi _{\alpha }$ ($\alpha =1,2$) and $p$ are complex-valued
functions. Given the potential $p(z,\overline{z})$ and solutions $\psi
_{\alpha },\varphi _{\alpha }$ of the system (\ref{e6}), one defines four
real-valued functions $X^{\mu }$ ($\mu =1,2,3,4$) as 
\begin{eqnarray}
X^{1}+iX^{2} &=&\int_{\Gamma }{\left( -\varphi _{1}\varphi _{2}dz^{\prime
}+\psi _{1}\psi _{2}d\overline{z}^{\prime }\right) }\;\,\,,  \nonumber \\
X^{1}-iX^{2} &=&\int_{\Gamma }{\left( \overline{\psi }_{1}\overline{\psi }%
_{2}dz^{\prime }-\overline{\varphi }_{1}\overline{\varphi }_{2}d\overline{z}%
^{\prime }\right) \quad }\;,  \nonumber \\
X^{3}+iX^{4} &=&\int_{\Gamma }{\left( \varphi _{1}\overline{\psi }%
_{2}dz^{\prime }+\psi _{1}\overline{\varphi }_{2}d\overline{z}^{\prime
}\right) }\;\quad ,  \nonumber \\
X^{3}-iX^{4} &=&\int_{\Gamma }{\left( \overline{\psi }_{1}\varphi
_{2}dz^{\prime }+\overline{\varphi }_{1}\psi _{2}d\overline{z}^{\prime
}\right) \quad \,\,,}  \label{e7}
\end{eqnarray}
where $\Gamma $ is a contour in the domain $G$ ($z\in G$). Formulae (\ref{e7}%
) give a conformal immersion of a surface with local coordinates $z,%
\overline{z}$ where $z=\xi _{1}+i\xi _{2}$ into $\Bbb{R}^{4}$.An advantage
of the GWR (\ref{e7}) is that many geometrical objects take a simple and
compact form. Namely, the induced metric is given by 
\begin{equation}
ds^{2}=u_{1}u_{2}\,dz\,d\overline{z}  \label{e8}
\end{equation}
where $u_{\alpha }=\left| \psi _{\alpha }\right| ^{2}+\left| \varphi
_{\alpha }\right| ^{2}$ ($\alpha =1,2$), the Gaussian curvature $K\,$and
normal curvature $K_{N}$ are of the form 
\begin{equation}
K=-\frac{2}{u_{1}u_{2}}\left[ \log \left( u_{1}u_{2}\right) \right] _{z%
\overline{z}}\quad \quad ,\quad \quad K_{N}=\frac{2}{u_{1}u_{2}}\left[ \log
\left( \frac{u_{2}}{u_{1}}\right) \right] _{z\overline{z}}  \label{e9}
\end{equation}
while the mean curvature vector $\overrightarrow{H}$ is 
\begin{eqnarray}
\overrightarrow{H} &=&\frac{2}{u_{1}u_{2}}\left[ Re\left( p\psi _{2}\varphi
_{1}+\overline{p}\psi _{1}\varphi _{2}\right) ,Im\left( p\psi _{2}\varphi
_{1}+\overline{p}\psi _{1}\varphi _{2}\right) ,\right.  \nonumber \\
&&\left. \;\;\;\;\;\;\;\;\;\;\;\;\;Re\left( p\varphi _{1}\overline{\varphi }%
_{2}-\overline{p}\psi _{1}\overline{\psi }_{2}\right) ,Im\left( p\overline{%
\psi }_{1}\psi _{2}-\overline{p}\overline{\varphi }_{1}\varphi _{2}\right)
\right] \;\quad .  \label{e10}
\end{eqnarray}
Using the formulae (\ref{e6})-(\ref{e10}), for the Nambu-Goto and the
Polyakov action one gets [\ref{re11},\ref{re12}] 
\begin{equation}
S_{N-G}=\mu _{0}\int u_{1}u_{2}\,d\xi _{1}\,d\xi _{2}\qquad ,  \label{e11}
\end{equation}
\begin{equation}
S_{P}=\frac{4}{\alpha _{0}}\int |p|^{2}d\xi _{1}\,d\xi _{2}\qquad \quad .
\label{e12}
\end{equation}
Then for the self-intersection number (or the total normal curvature, or the
first Chern number $c_{1}(\Sigma )$ ), one gets the following expression 
\begin{equation}
I=\frac{1}{2}c_{1}=\frac{i}{8\pi }\int \int_{G}\left( \log \frac{u_{2}}{u_{1}%
}\right) _{z\,\overline{z}}\,dz\wedge d\overline{z}=\frac{i}{8\pi }%
\int_{\partial G}\left( \log \frac{u_{2}}{u_{1}}\right) _{z\,}\,dz\,
\label{e13}
\end{equation}
where $\partial G$ denotes the boundary of the domain $G$. For the tensors $%
t_{\pm }^{\mu \nu }\stackrel{def}{=}t^{\mu \nu }\pm t^{*\mu \nu }\,$, one
obtains 
\begin{eqnarray}
t_{+}^{12} &=&t_{+}^{34}=\frac{|\varphi _{1}|^{2}-|\psi _{1}|^{2}}{|\varphi
_{1}|^{2}+|\psi _{1}|^{2}}\quad \quad ,  \nonumber \\
t_{+}^{13} &=&t_{+}^{42}=i\frac{\psi _{1}\varphi _{1}-\overline{\psi }_{1}%
\overline{\varphi }_{1}}{|\varphi _{1}|^{2}+|\psi _{1}|^{2}}\quad ,
\label{e14} \\
t_{+}^{14} &=&t_{+}^{23}=\frac{\psi _{1}\varphi _{1}+\overline{\psi }_{1}%
\overline{\varphi }_{1}}{|\varphi _{1}|^{2}+|\psi _{1}|^{2}}  \nonumber
\end{eqnarray}
and 
\begin{eqnarray}
t_{-}^{12} &=&t_{-}^{43}=\frac{|\varphi _{2}|^{2}-|\psi _{2}|^{2}}{|\varphi
_{2}|^{2}+|\psi _{2}|^{2}}\quad \quad ,  \nonumber \\
t_{-}^{13} &=&t_{-}^{24}=i\frac{\psi _{2}\varphi _{2}-\overline{\psi }_{2}%
\overline{\varphi }_{2}}{|\varphi _{2}|^{2}+|\psi _{2}|^{2}}\quad ,
\label{e15} \\
t_{-}^{14} &=&t_{-}^{32}=-\frac{\psi _{2}\varphi _{2}+\overline{\psi }_{2}%
\overline{\varphi }_{2}}{|\varphi _{2}|^{2}+|\psi _{2}|^{2}}\quad . 
\nonumber
\end{eqnarray}
Now let us consider self-dual and anti-self-dual configurations. They are
given by equation 
\begin{equation}
t_{\mu \nu }^{\mp }=c_{\mu \nu }^{\mp }  \label{e16}
\end{equation}
where $c_{\mu \nu }^{\pm }$ are constant tensors.

In the self-dual case, using (\ref{e15}), one gets 
\begin{equation}
a\varphi _{2}=b\overline{\psi }_{2}  \label{e17}
\end{equation}
where $a$ and $b$ are arbitrary constants. Effectively one has, obviously,
only one arbitrary constant.

The constraint (\ref{e17}) is compatible with the system (\ref{e6}) if $p=0$%
. So, in virtue of (\ref{e10}), the self-dual configurations are given by
minimal surfaces in $\Bbb{R}^{4}$. This result is not new (see \emph{e.g.} [%
\ref{re9}]). Further, using the GWR (\ref{e17}), one obtains 
\begin{equation}
\overrightarrow{X}_{zz}\cdot \overrightarrow{X}_{zz}=\left( \overline{\psi }%
_{1}\varphi _{1z}-\overline{\psi }_{1z}\varphi _{1}\right) \left( \overline{%
\psi }_{2z}\varphi _{2}-\overline{\psi }_{2}\varphi _{2z}\right) \quad .
\label{e18}
\end{equation}
Taking into account the constraint (\ref{e17}), one concludes that 
\begin{equation}
\left( \overrightarrow{X}_{zz}\right) ^{2}=0\quad .  \label{e19}
\end{equation}
The constraint (\ref{e19}) defines the, so-called, superminimal immersion
when at each point of a surface its curvature ellipse is a circle [\ref{re13}%
].

So instantons in the theory (\ref{e1}) are given by surfaces superminimally
immersed into $\Bbb{R}^{4}$.

Similarly, in the anti-self-dual case the expressions (\ref{e14}) give the
constraint 
\begin{equation}
c\varphi _{1}=d\overline{\psi }_{1}  \label{e20}
\end{equation}
where $c\,\,$and $d$ are arbitrary constants. Again, this constraint implies
that $p=0$ and $\left( \overrightarrow{X}_{zz}\right) ^{2}=0$.

So, both instantons and anti-instantons of the string theory in $\Bbb{R}^{4}$
governed by the action (\ref{e1}) are given by the formulae (\ref{e6}), (\ref
{e7}) with $p=0$ and the constraints (\ref{e17}) or (\ref{e20}),
respectively.

Superminimal immersions in $\Bbb{R}^{4}$ have attracted essential interest
in differential geometry (see \emph{e.g.} [\ref{re13},\ref{re14}]). An
advantage of the GWR in this case is that it allows to obtain simple
explicit formulae for such immersions. Namely, for the self-dual case the
formulae (\ref{e7}) imply 
\begin{equation}
\begin{array}{l}
a\left( X^{1}+iX^{2}\right) _{z}=-b\,\varphi _{1}\overline{\psi }_{2}\quad
\quad , \\ 
\left( X^{3}+iX^{4}\right) _{z}=\varphi _{1}\overline{\psi }_{2}\quad \quad
\quad \quad ,
\end{array}
\qquad 
\begin{array}{l}
\left( X^{1}+iX^{2}\right) _{\overline{z}}=\psi _{1}\psi _{2}\quad \quad
\quad , \\ 
\overline{a}\left( X^{3}+iX^{4}\right) _{\overline{z}}=\overline{b\,}\psi
_{1}\psi _{2}\quad \quad
\end{array}
\label{e21}
\end{equation}
and for the anti-self-dual case one has 
\begin{equation}
\begin{array}{l}
\left( X^{1}-iX^{2}\right) _{z}=\overline{\psi }_{1}\overline{\psi }%
_{2}\quad \quad , \\ 
c\left( X^{3}+iX^{4}\right) _{z}=d\,\overline{\psi }_{1}\overline{\psi }%
_{2}\quad ,
\end{array}
\qquad 
\begin{array}{l}
\overline{c}\left( X^{1}-iX^{2}\right) _{\overline{z}}=-\overline{d\,}\psi
_{1}\overline{\varphi }_{2}\quad \quad , \\ 
\left( X^{3}+iX^{4}\right) _{\overline{z}}=\psi _{1}\overline{\varphi }%
_{2}\quad \quad \quad \quad .
\end{array}
\label{e22}
\end{equation}
where $\psi _{1},\psi _{2}$ are arbitrary anti-holomorphic functions and $%
\varphi _{1},\varphi _{2}\,$are arbitrary holomorphic functions.

The relations (\ref{e21}) give 
\begin{eqnarray}
a\left( X^{1}+iX^{2}\right) +b\left( X^{3}+iX^{4}\right) &=&A\left( 
\overline{z}\right) \quad ,  \nonumber \\
\overline{b}\left( X^{1}+iX^{2}\right) -\overline{a}\left(
X^{3}+iX^{4}\right) &=&B\left( z\right) \quad  \label{e23}
\end{eqnarray}
where $A\left( \overline{z}\right) $ and $B\left( z\right) $ are arbitrary
anti-holomorphic and holomorphic functions, respectively. In the case (\ref
{e22}) one gets 
\begin{eqnarray}
d\left( X^{1}-iX^{2}\right) -c\left( X^{3}+iX^{4}\right) &=&C\left( 
\overline{z}\right) \quad ,  \nonumber \\
\overline{c}\left( X^{1}-iX^{2}\right) +\overline{d}\left(
X^{3}+iX^{4}\right) &=&D\left( z\right) \quad  \label{e24}
\end{eqnarray}
where $C\left( \overline{z}\right) $ and $D\left( z\right) $ are arbitrary
anti-holomorphic and holomorphic functions.

Hence, for instantons 
\begin{equation}
\begin{array}{l}
X^{1}+iX^{2}=\frac{1}{|a|^{2}+|b|^{2}}\,\left[ \overline{a}A\left( \overline{%
z}\right) +bB\left( z\right) \right] \quad , \\ 
\\ 
X^{3}+iX^{4}=\frac{1}{|a|^{2}+|b|^{2}}\,\left[ \overline{b}A\left( \overline{%
z}\right) -aB\left( z\right) \right]
\end{array}
\label{e25}
\end{equation}
and for anti-instantons 
\begin{equation}
\begin{array}{l}
X^{1}-iX^{2}=\frac{1}{|c|^{2}+|d|^{2}}\,\left[ \overline{d}C\left( \overline{%
z}\right) +cD\left( z\right) \right] \quad , \\ 
\\ 
X^{3}+iX^{4}=\frac{1}{|c|^{2}+|d|^{2}}\,\left[ -\overline{c}C\left( 
\overline{z}\right) +dD\left( z\right) \right] \quad .
\end{array}
\label{e26}
\end{equation}
Thus, instantons and anti-instantons are given by explicit formulae (\ref
{e25}) and (\ref{e26}) where $A,B,C,D$ are arbitrary functions and $a,b,c,d$
are arbitrary constants. Particular cases of these formulae corresponding to 
$a=0$ or $b=0$, $c=0$ or $d=0$ $\,$have been found, in fact, in the paper [%
\ref{re7}] where some concrete examples have been also presented.

Now we will consider some other interesting configurations different from
instantons.

The developable surfaces (surfaces for which the Gaussian curvature $K=0$)
are the simplest of them. They are given by the GWR (\ref{e7}) where $\psi
_{\alpha },\varphi _{\alpha }$ ($\alpha =1,2$) obey the system (\ref{e6})
with the additional constraint 
\begin{equation}
\left( \left| \psi _{1}\right| ^{2}+\left| \varphi _{1}\right| ^{2}\right)
\left( \left| \psi _{2}\right| ^{2}+\left| \varphi _{2}\right| ^{2}\right)
=\left| A(z)\right| ^{2}  \label{e27}
\end{equation}
where $A(z)$ is an arbitrary holomorphic function.

Another class of surfaces, the so-called surfaces with flat normal bundle,
correspond to the vanishing normal curvature $K_{N}=0$. In virtue of (\ref
{e9}), they are generated by the GWR (\ref{e6}), (\ref{e7}) with the
additional constraint 
\begin{equation}
\left( \left| \psi _{2}\right| ^{2}+\left| \varphi _{2}\right| ^{2}\right)
=\left| A(z)\right| ^{2}\left( \left| \psi _{1}\right| ^{2}+\left| \varphi
_{1}\right| ^{2}\right)   \label{e28}
\end{equation}
where $A(z)$ is an arbitrary holomorphic function. Surfaces with flat normal
bundle represent themselves an important class of surfaces in $\Bbb{R}^{4}$
(see \emph{e.g.} [\ref{re15},\ref{re16}]).

Choosing $A=1$, differentiating (\ref{e28}) with respect to $z$ and using (%
\ref{e6}), one gets 
\begin{equation}
\psi _{1\overline{z}}\overline{\psi }_{1}+\varphi _{1}\overline{\varphi }_{1%
\overline{z}}=\psi _{2\overline{z}}\overline{\psi }_{2}+\varphi _{2}%
\overline{\varphi }_{2\overline{z}}\quad \quad .  \label{e29}
\end{equation}
This condition is satisfied, in particular, if there are functions $q$ and $%
\widetilde{q}$ such that 
\begin{equation}
\begin{array}{l}
\psi _{1\overline{z}}=q\varphi _{1}\quad \quad , \\ 
\varphi _{1z}=-\overline{q}\psi _{1}\quad \,\,,
\end{array}
\qquad 
\begin{array}{l}
\psi _{2\overline{z}}=\overline{\widetilde{q}}\varphi _{2}\quad \quad , \\ 
\varphi _{2z}=-\widetilde{q}\psi _{2}\qquad .
\end{array}
\label{e30}
\end{equation}
Equations (\ref{e6}) and (\ref{e30}) imply 
\begin{equation}
\begin{array}{l}
q\psi _{1z}-p\psi _{1\overline{z}}=0\quad \quad , \\ 
\overline{\widetilde{q}}\psi _{2z}-\overline{p}\psi _{2\overline{z}}=0\quad
\quad ,
\end{array}
\qquad 
\begin{array}{l}
\overline{p}\varphi _{1z}-\overline{q}\varphi _{1\overline{z}}=0\quad \quad ,
\\ 
\overline{p}\varphi _{2z}-\overline{\widetilde{q}}\varphi _{2\overline{z}%
}=0\qquad .
\end{array}
\label{e31}
\end{equation}
Equations (\ref{e31}) are of the Beltrami type. The theory of such type of
equations is well-developed (see \emph{e.g.} [\ref{re17}]). So, one has the
class of surfaces in $\Bbb{R}^{4}$ with flat normal bundle generated by the
GWR formulae (\ref{e7}) where $\psi _{\alpha }$ and $\varphi _{\alpha }$ ($%
\alpha =1,2$) are defined by equations (\ref{e31}).

Surfaces with constant (non-vanishing) mean curvature $\overrightarrow{H}^{2}
$ provide another class of surfaces. In virtue of the formula $%
\overrightarrow{H}^{2}=4\frac{|p|^{2}}{u_{1}u_{2}}$ these surfaces are given
by the GWR (\ref{e7}) where $\psi _{\alpha }$ and $\varphi _{\alpha }$
satisfy the system of equations 
\begin{eqnarray}
\psi _{\alpha z} &=&\frac{1}{2}\exp (i\theta _{\alpha })\,\sqrt{\vec{H}%
^{2}\left( |\psi _{1}|^{2}+|\varphi _{1}|^{2}\right) \left( |\psi
_{2}|^{2}+|\varphi _{2}|^{2}\right) }\;\varphi _{\alpha }\quad ,  \nonumber
\\
\varphi _{\alpha \overline{z}} &=&-\frac{1}{2}\exp (-i\theta _{\alpha })\,%
\sqrt{\vec{H}^{2}\left( |\psi _{1}|^{2}+|\varphi _{1}|^{2}\right) \left(
|\psi _{2}|^{2}+|\varphi _{2}|^{2}\right) }\;\psi _{\alpha }\quad ,\quad
\alpha =1,2  \nonumber \\
&&  \label{e32}
\end{eqnarray}
where $\theta _{1}(z,\overline{z})=-\theta _{2}(z,\overline{z})$ is an
arbitrary function. A subclass of such surfaces, the so-called surfaces with
the harmonic Gauss map, is associated with the additional constraints 
\begin{equation}
p=a\,(\,|\psi _{1}|^{2}+|\varphi _{1}|^{2})=b(\,|\psi _{2}|^{2}+|\varphi
_{2}|^{2})\quad   \label{e33}
\end{equation}
where $a$ is an arbitrary constant and $b=\frac{\vec{H}^{2}}{4\overline{a}}$%
. So, surfaces with harmonic Gauss map are given by (\ref{e6}) where $\psi
_{\alpha }$ and $\varphi _{\alpha }$ obey the system 
\begin{eqnarray}
\psi _{1z} &=&\,a\,(\,|\psi _{1}|^{2}+|\varphi _{1}|^{2})\;\varphi _{1}\quad
\,\,\,,\quad \quad \psi _{2z}=\,\overline{a}\,(\,|\psi _{1}|^{2}+|\varphi
_{1}|^{2})\;\varphi _{2}\quad \quad ,  \nonumber \\
\varphi _{1\overline{z}} &=&-\overline{a}\,(\,|\psi _{1}|^{2}+|\varphi
_{1}|^{2})\;\psi _{1}\quad ,\quad \quad \varphi _{2\overline{z}%
}=-a\,(\,|\psi _{1}|^{2}+|\varphi _{1}|^{2})\;\psi _{2}\quad \,\,.
\label{e34}
\end{eqnarray}

\textbf{Acknowledgment.} The authors are grateful to Dr. J. Pawe\l czyk for
attracting their attention to papers [\ref{re8}-\ref{re10}].

\begin{centerline}
\textbf{REFERENCES}
\end{centerline}

\begin{enumerate}
\item  \label{re1} D.J. Gross, T. Piran and S. Weinberg (Eds.), \textit{Two
dimensional quantum gravity and random surfaces}, World Scientific,
Singapore, 1992.

\item  \label{re2}F. David, P. Ginsparg and Y. Zinn-Justin (Eds.), \textit{%
Fluctuating geometries in Statistical Mechanics and Field Theory}, Elsevier
Science, Amsterdam, 1996.

\item  \label{re3}A.M. Polyakov, Nucl. Phys. \textbf{B268}, 406 (1986).

\item  \label{re4}A.P. Balachandran, F. Lizzi and G Sparano, Nucl. Phys. 
\textbf{B263}, 608 (1986).

\item  \label{re5}P.O. Mazur and V.P. Nair, Nucl. Phys. \textbf{B284}, 146
(1986).

\item  \label{re6}J.F. Wheater, Phys. Lett \textbf{B208}, 244 (1988)

\item  \label{re7}G.D. Robertson, Phys. Lett \textbf{B226}, 244 (1989).

\item  \label{re8}J. Pawe\l czyk, Phys. Rev. Lett. \textbf{74}, 3924 (1995).

\item  \label{re9}J. Pawe\l czyk, Phys. Lett. \textbf{B387}, 287, (1996).

\item  \label{re10}J. Pawe\l czyk, Nucl. Phys. \textbf{B491}, 515 (1997).

\item  \label{re11}B.G. Konopelchenko, \textit{Weierstrass representation
for surfaces in 4D spaces and their integrable deformations via the DS
hierarchy}, preprint, math.DG/9807129, (1998).

\item  \label{re12} B.G. Konopelchenko and G. Landolfi, \textit{Induced
surfaces and their integrable dynamics. II. Generalized Weierstrass
representations in 4D spaces and deformations via DS hierarchy}, math.DG/
9810138, (1998).

\item  \label{re13}R.L. Bryant, J. Diff. Geom. \textbf{17}, 455 (1982).

\item  \label{re14} T. Friedrich, Ann. Glob. Analysis and Geometry, \textbf{2%
}, 257 (1984).

\item  \label{re15}C.-L. Terng, Math. Ann., \textbf{277}, 95 (1987).

\item  \label{re16}E.V. Ferapontov, \textit{Surfaces with flat normal
bundle: an explicit construction}, preprint SFB288, N.320, TU-Berlin (1998).

\item  \label{re17}I.N.Vekua, \textit{Generalized analytic functions},
Nauka, Moscow, 1988.
\end{enumerate}

\end{document}